\documentclass[
aip,jap, jmp,amsmath,amssymb,
preprint,%
reprint,%
numerical,two-column, floatfix]{revtex4-1}
\usepackage{lineno,hyperref}
\usepackage{graphicx}
\usepackage{amssymb}
\usepackage{lineno}
\usepackage[T1]{fontenc}   
\usepackage{bm}
\usepackage[utf8]{inputenc}
\usepackage{amsmath}
\usepackage{amsfonts}
\usepackage{wasysym}
\usepackage{setspace}
\usepackage{xfrac}
\usepackage{xcolor}
\usepackage{epstopdf}
\usepackage{miller}
\modulolinenumbers[5]

\date{March 2020}

\begin{document}
\title{Transmission Helium Ion Microscopy of Graphene}	

\email{kavanagh@sfu.ca.}
\author{Karen L. Kavanagh}
\author{Aleksei Bunevich}

\affiliation{Physics Department, Simon Fraser University, 8888 University Dr., Burnaby, BC, Canada V5A 1S6}
\author{Mallikarjuna Rao Motapothula}
\affiliation{Department of Physics, SRM University AP, Amaravati, Andhra Pradesh 522-502, India
}

\begin{abstract}
$\mathbf{Abstract}$: We compare transmission HIM to transmission electron microscopy (TEM) of graphene support films.  We  present spot transmission patterns that compare with SRIM predictions, and show examples of scanning He$^+$ transmission images, based on integrated camera intensity. We also consider the potential for coherent HIM scattering.  
\end{abstract}
\maketitle
\section{Introduction}

Accelerated He ions have been tremendously successful at atomic analysis, using a wide range of beam energies and techniques. Elastic  collisions, the basis for Rutherford backscattering (RBS), measures atomic composition and depth profiles via known nuclear scattering cross-sections and rates of energy loss.\cite{Chu1978} Ion channeling, where ions travel preferentially between atomic planes, is exploited for many crystalline defect questions.\cite{Vantomme} Both techniques rely on stable accelerators, minimal beam divergence via collimation slits with long beam lines, and sufficient energy resolution in the detectors (15 keV in the case of typical Si diodes). Channeling is understood to occur via a reduced scattering cross-section occurring for shallow scattering angles with respect to atomic planes. The degree of spatial coherence (fraction of the He beam that is in phase given sufficient temporal coherence) is important for the detection of diffraction. Surface elastic scattering with diffraction patterns in ultra high vaccum has been reported for thermal and fast He ions and atom beams.\cite{WINTER2011169}, and is another confirmation of the de Broglie relationship. 

The He ion microscope (HIM) is a commercial, focussed ion beam (FIB) instrument (Zeiss, Nanofab) well established for nanoscale milling, and ion-induced, secondary electron imaging (SEI).\cite{Hlawacektext}  The He$^+$ ions field emit from a single W atom, part of a surface trimer on the W tip held at an accelerating voltage of 10 kV to  40 kV. Ion beam current  (0.01 to 100 pA) on the sample is controlled by apertures and He gas pressure, and focussed onto the sample via 2 electrostatic lenses. The higher mass of He results in a smaller de Broglie wavelength (80 fm) and a larger depth of focus, compared to electrons in typical scanning electron microscopes (EM). Similar to the thermal He beam experiments, the small source volume of the HIM means that the spatial coherence of this ion beam is expected to be much greater than those of the classical, collimated MeV beams. Plus, the higher energy loss at these lower keV energies may be more useful for analysis of thin samples with low mass, than is previously reported for electron beams. However, electron uptake and reionization of the He ions during transit through a sample  complicate these interactions.\cite{DALFONSO201318, Primetzhofer_PRL_2020}   

We have added a digital camera below the sample stage in our HIM.\cite{Kavanagh2017} This has allowed the direct detection of transmitted He ions and neutral atoms that pass through thin samples. The camera is currently a modified x-ray detector consisting of an array of Si diode pixels (each 55 µm square) without surface layers such as minilens or optical filters (Modupix, Advacam). Given that the HIM beam energies can be up to 40 keV, the He penetrates into the Si pixels to a maximum average range of 350 nm, generating a current pulse proportional to its energy. We have demonstrated focused He ion channeling through single crystalline Si (100)  membranes (50 nm), measuring [011] critical incident angles of 1º at 35 keV, and beam steering of 2º.\cite{Kavanagh2018doi:10.1116/1.5020667} 

In this paper we compare transmission HIM (THIM) to transmission EM (TEM) using single and multilayer graphene support grids to gain insights into what is required to perhaps harness the expected coherence of the HIM beam.

\section{Experimental}
The samples consisted of graphene (1 or 6-8 layers) supported on Cu grids (commercially available from Ted Pella). The structural properties were investigated by field-emission scanning and transmission EM using a field-emission gun system (operating at 200 keV), and a HIM (operating at 30 keV). Both beams are known to damage graphene rapidly so care was taken to minimize exposures.

Bright field (BF) TEM images were obtained by inserting an aperture (objective) around the centre beam to enhance contrast from scattered electrons. Low dose images were obtained by rapid displacement of the beam from the focussing spot to lateral areas where an image was obtained via a 0.5 sec exposure with a CCD camera. Associated selected area diffraction patterns are obtained by lenses underneath the sample that focus on the diffraction plane from a region of sample limited by a second aperture below the sample (selected area aperture). 

HIM SEI images were obtained by scanning the focussed He beam at rates where no changes were observed in the film. The total incident HIM ion beam current, detected by deflecting the beam onto a Faraday cup associated with blanking plates in the column, was typically less than 0.1 pA. The total secondary electron emission as a function of spot position results in an SEI image. THIM images were collected  using the camera installed at the bottom of our HIM chamber, 20 cm below the sample.\cite{Kavanagh2017} The camera has a total area 14 x 14 mm$^2$, giving a maximum scattering angle of $\pm$70 mrad. Both single spot and scanned FIB exposures were obtained.  The current detected by the camera depends on electron-hole pair generation that linearly decreases with ion energy.\cite{Kavanagh2018doi:10.1116/1.5020667} 

With our 2D camera, we generated scanning transmission HIM images by integrating the signal at large scattering angles, as a function of spot location. Like annular dark-field STEM images, the signal is sensitive to the mass of the scattering sample atom. The ion beam scanning frequency was correlated with that of the camera exposure and data processing  (frame speed, dead-time) through an in-house Matlab routine. Through application of either the HIM controls or the associated ion-beam lithography software (Fibics), the beam was stopped or rastered over an area at a controlled dose rate. 

\section{Results}
Figure 1 shows typical TEM bright-field (BF) images and diffraction patterns of a  region of the graphene samples (a) 1 layer and (b) 6-8 layers. The image contrast is variable with evidence of foreign material. Nevertheless, the dominant diffraction patterns in both cases are hexagonal with uniform spot intensities. These spots are due to diffraction from thin planes almost parallel to the beam, with Bragg angles for 200 keV electrons of only 12 mrad. Uniform spot intensities are expected for single layer graphene but not for bilayers,  multilayer graphene or graphite where the intensity of  the second row spots ($\{$1120$\}$) are expected to be greater than the first row ($\{$1100$\}$).\cite{GrapheneTEM_PARK2010797} We also see a smaller polycrystalline ring in the 6-8 graphene sample indicating likely a second phase contamination. Higher magnification does resolve atomic lattice fringes with spacings of 0.21 nm, but there is much evidence of random amorphous contamination. 

The same two samples were imaged using SEI with the HIM as shown in Figure 2 (a) and (b). These images show one entire opening in each Cu grid where the graphene support material should be. Overall, the single layer sample in (a) is darker than that of the thicker 6-8 layer sample in (b). Graphene wrinkles are visible in (a) commonly observed in such films. In the thicker sample in (b), higher intensity white SEI contrast can be due to less channeling.\cite{Hlawacek2012} The black regions are holes in the film.

The inset in (b) shows an example of a scanning transmission HIM image that confirms the identification of each region. The holes are now the brightest while the white regions are the darkest where the least transmission occurred.  The resolution is a function of spot size approximately 1 nm. 

An example of a focussed spot transmission scattering pattern is shown in Fig. 3 (a) for a 6-8 layer graphene sample compared with a SRIM\cite{SRIM}  simulation in (b) for transmission through similar mass, amorphous carbon (2 nm). The experimental scattering pattern from the entire camera (1.4 x 1.4 cm$^2$) is visible. Black regions are dead pixels or where the camera is blocked by the stage. The transmitted He pattern is rotationally symmetric like the SRIM simulation with transmitted ions detected out to the maximum range of the camera. To more accurately compare them, we have computed the probability as a function of scattering angle for this data and for single layer graphene sample, and compared them with SRIM predictions. The results are plotted in Fig. 4. The experimental profiles are well predicted by SRIM with maximum scattering angles between 5 and 10 mrad for the graphene samples that compare with the predictions for 2.5 nm $\alpha$-carbon. 

Finally, it is apparent in the image in (a) that the camera has developed many dead pixels likely from He bubbles forming within the Si lattice when we exceed an implant dose greater than 10$^{17}$cm$^-3$.\cite{Livengood} We are investigating whether annealing the camera during or after operation will help to avoid this problem. 

\section{Discussion}
The STHIM results shows us that we can  use the transmitted intensity as a function of position to develop scanning THIM. This technique would be more sensitive to thickness variations than the electron variety that uses annular dark field detectors. STHIM would be especially useful for low mass elements. 

TEM of graphene easily produces diffraction patterns indicating a beam with sufficient coherence. EM relies on an electron source with small emission volumes and narrow emission bandwidths combined with high current. These characteristics determine the degree of spatial and temporal  coherence of the beam, respectively. Temporal coherence is associated with the width of the beam energy or wavelengths, a function primarily of the stability of the accelerator power supply. Thus, the energy or wavelength of each incident particle will differ by perhaps 1 ppm in the latest instruments. Spatial coherence describes the volume of sample that experiences the same phase in the incident radiation given the best temporal coherence. A truly planar wave does not exist. Smaller sources provide  higher spatial coherence given apertures and geometries that subtend small angles. Increased coherence from field-emission sources is recognized by the strength of diffraction from elastic scattering, the contrast ratio in holograms, and the information limit of atomic lattice images. 

The best EM resolution today ($\approx$ 50 pm) is via electron transmission through suitably thinned samples, limited not by the electron wavelength (2 pm at 300 keV) but by lens aberrations. (Thin samples  are necessary since energy losses broaden the beam energy and quality of focus.) Diffraction spot separations are related to the wavelength of the electron, following Bragg's law, but lens aberrations degrade the quality of focussing required to produce uniform beams and magnifications.
With the small He ion source and column length of the HIM, we expect the spatial coherence of the HIM beam to be sufficient, but our camera and scattering geometry do not have the resolution required to detect HIM diffraction if occurring. 

Considering the $\{$1100$\}$  graphene spots, given their planar spacing of 0.21 nm, the expected first order Bragg angle is 0.2 mrad, 100 times smaller than for the TEM case. This scattering angle corresponds to a distance of 50 $\mu m$ at the camera, or one pixel from the incident beam path. We can focus the beam spot entirely within 1 of our pixels but resolving a potential diffraction pattern will require a larger camera length, or a higher density of smaller pixels.

\section{Declaration of Competing Interests}
The authors declare that they have no known competing financial interests or personal relationships that could have appeared to influence the work reported in this paper.
\section{Acknowledgments}
We are grateful for support from the Natural Science and Engineering Council of Canada (NSERC RGPIN-2019-05086).

\begin{figure*}
	\includegraphics[width=0.5\columnwidth]{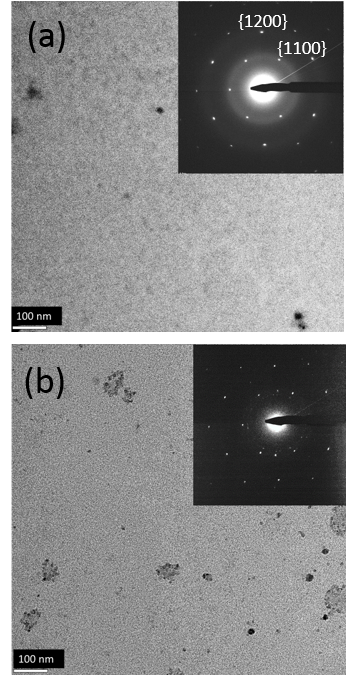}
	\caption{TEM bright field images and selected area diffraction patterns from graphene support grids (a) single layer and (b) 6-8 layers.}
\end{figure*}

\begin{figure*}
	\includegraphics[width=0.5\columnwidth]{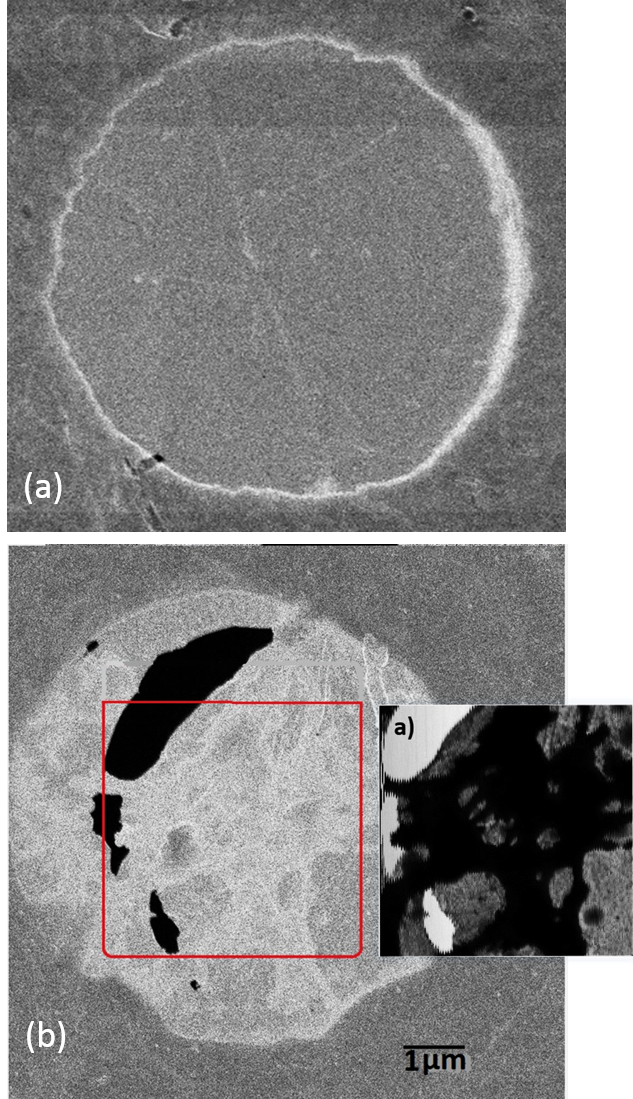}
	\caption{HIM secondary electron emission images from graphene support grids (a) single layer and (b) 6-8 layers. The insert in (b) is a image obtained by scanning transmission HIM}
\end{figure*}

\begin{figure*}
	\includegraphics[width=0.5\columnwidth]{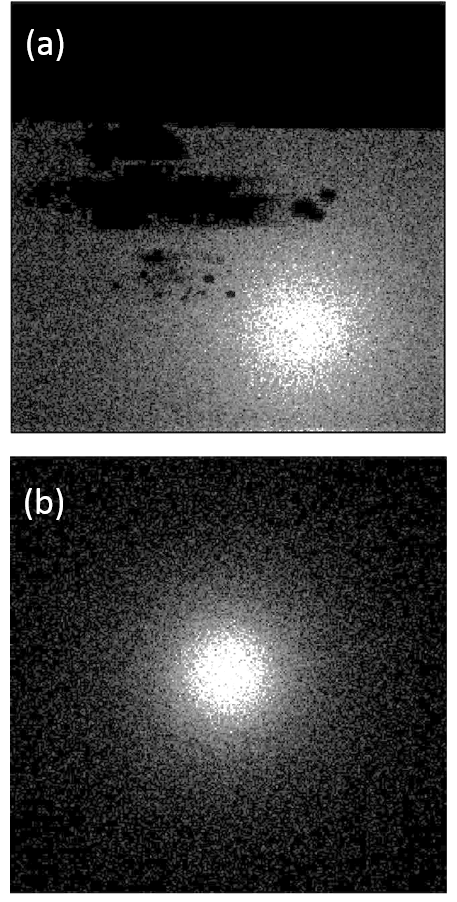}
	\caption{Spot exposure (a) showing a transmission scattering pattern for 35 keV He$^+$ ions into graphene 6-8 layers. (b) Simulation of the same exposure for amorphous carbon (2.5 nm). Shown is the full camera (1.4 x 1.4 cm$^2$ area) with 256 x 256 pixels each 55 $\mu$m$^2$.}
\end{figure*}

\begin{figure*}
	\includegraphics[width=\columnwidth]{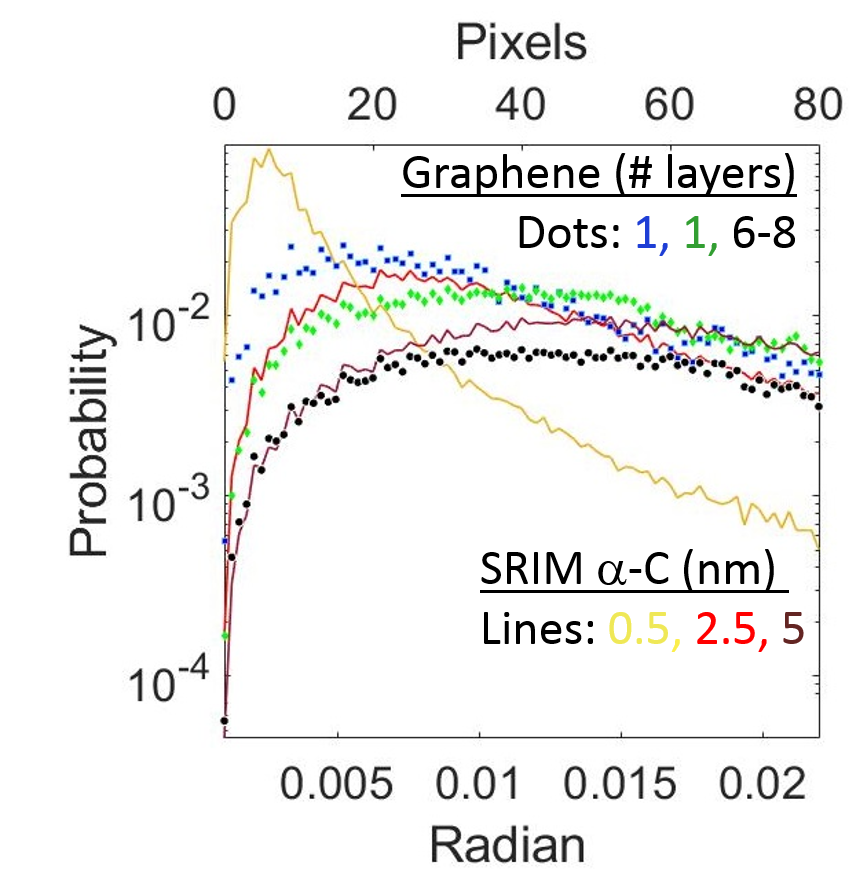}
	\caption{Scattering probability as a function of scattering angle for spot exposures of graphene samples compared with SRIM simulations of amorphous carbon layers of equivalent mass. The beam energy was 35 keV.  }
\end{figure*}

\bibliography{Arxiv}
\end{document}